# Software system rationalisation: How to get better outcomes through stronger user engagement.

## Authors

Richard E. Shute*[1] and Nick Lynch[2] (Curlew Research[3], Woburn Sands, UK)

## Abstract

As businesses get more sizable and more mature they now, inevitably accrete more and more software systems. This estate expansion leads not only to greater complexity and expense for the enterprise, but also to fragmentation, inconsistency and siloing of business processes. Because platform rationalisation and system decommissioning never happens spontaneously, a perennial problem for the enterprise then becomes how to simplify their corporate software platforms. Recently, Curlew Research personnel were involved in a software rationalisation program within a large global life sciences company and this paper describes an approach to decommissioning which we developed as part of that project, and which we feel could be of use more widely to help with objective more user-centric system rationalisation. The method derives from a model developed by Noriaki Kano et al to help with determining customer satisfaction and loyalty, and the prioritisation of new, additional functionality, features or "products", for example when looking to enhance software applications. Using a blueprint process for rationalisation, the Curlew-Kano method enables each application to be placed efficiently and objectively into one of four categories - **Retain; Review; Remove; Research** - thus allowing the enterprise to identify and prioritise quickly those systems which warrant further investigation as part of a decommissioning activity. The key difference of the Curlew-Kano method compared to other application rationalisation methodologies is the fundamental involvement of users in the identification of systems more suitable for rationalisation and possible decommissioning. In our view involving users more fully in system rationalisation leads to better outcomes for the enterprise.

## 1. Introduction

**A perennial problem for the enterprise.**

As businesses get older, more mature and bigger they inevitably accrete more and more software systems, particularly now that "IT" has become as critical to businesses as electricity and water. This estate expansion leads not only to greater complexity and expense for the enterprise, but also to fragmentation, inconsistency and siloing of business processes. The big challenge then becomes how to rationalise and simplify the global enterprise software platforms, because system decommissioning never happens spontaneously. There is always someone who wants to retain 10-year old Application 'X', even though no-one else in the company uses it, because it is "absolutely critical" to what they do and you just CAN'T get rid of it! And there is always someone else who says the business needs brand new Application 'Y' to support innovation in their area because there is no system in the current platform that does quite what they need it to do.

### High tension

As a consequence of this drive towards an ever-expanding software universe, there is a constant tension and conflict in the enterprise between two stakeholder groups with two very different goals. On the one hand, you have the "Contractioners" – the software licence and support budget holders along with the business performance personnel, who are looking, respectively, to drive down costs and to drive up more consistent business working by keeping the software estate as compact as possible. Then on the other hand you have the "Expanders"; these are the innovators and the Luddites (ironically!) who want either to add new shiny software, or to keep the good old systems the same. How then, when the costs and the business process inconsistencies get too large, do enterprises deal with these two competing drivers of expansion and contraction (Figure 1)?

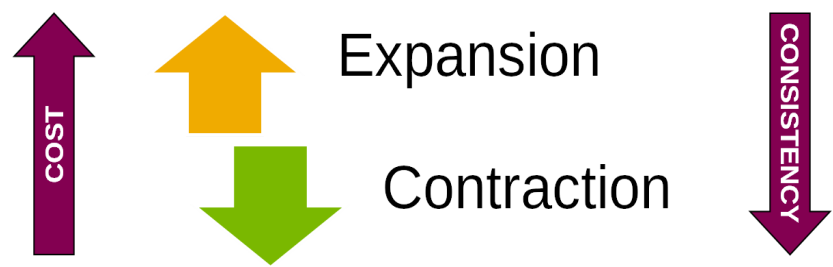

**Figure 1**: The software estate expansion-contraction conundrum

Commonly, the organisation runs a business change, software rationalisation program, and often they hand such a potentially contentious activity to a neutral organisation to run the change project. This makes a lot of sense as an independent, disinterested party should be (hopefully) unbiased in its approach, personnel, scope and reach, and in its final recommendations. But how can such an activity be truly objective in an area that is so prone to rampant subjectivity?

Recently, Curlew Research personnel were involved in a software rationalisation program within a large global life sciences company and we would like to share with you an approach to decommissioning that we developed as part of that project, and which we feel could be of use more widely to help with objective system decommissioning. The approach we developed was based on the Kano Model for determining customer satisfaction and loyalty. This model was developed by Noriaki Kano (Kano et al, 1984) in the early 1980's to help with the prioritisation of new, additional functionality, features or "products", for example when looking to enhance software applications.

## 2. The Curlew-Kano Method

The Kano Model for customer satisfaction and feature prioritisation is founded upon the following premise: that a customer's satisfaction with a product's potential new features will depend on the level of functionality that is provided; both how much (quantity) and how well (quality) those features are implemented.

Furthermore, features or functionality can be classified into four categories:-

1. **Must-Be**
    - Functionality *expected* by customers. If the product doesn't have this feature, it will be considered to be incomplete or just plain bad.
2. **Performance**
    - Increase in this functionality leads to increased satisfaction.
3. **Attractive**
    - Unexpected functionality and "wow" features, which, when presented, causes a positive reaction.
4. **Indifferent**
    - Functionality whose presence (or absence) doesn't make a real difference in the reaction to the product.

These categories can be plotted on a "Satisfaction-Functionality" graph (Figure 2) (Zacarias, 2022):-

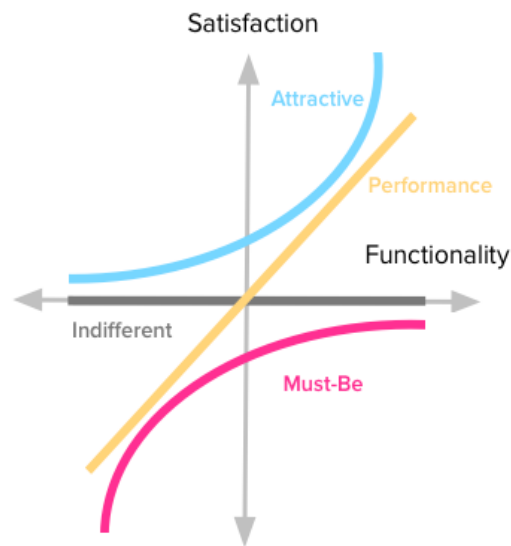

Figure 2: The Kano Model Satisfaction-Functionality Graph

We are not going to delve more here into the Kano Model in its classical application, so if you want to know more about its use in feature prioritisation, then we suggest reading these two articles (Zacarias, 2022; Product.pm, 2022).

### I Can't Get No Satisfaction

What attracted us to the potential of the Kano Model for system rationalisation was the fact that you can determine how customers rate the value of a feature through a simple two-question survey with, for each question, four possible answers. The first question focuses on a person's satisfaction with a feature and is more subjective. The second focuses on its functionality and usefulness and is therefore more objective. Switching the more traditional Kano Model questions to focus on systems and tools rather than individual features of those systems, we generated the following question and answer set:-

*How do you feel about System/Tool 'X' now?*
1. I like it.
2. I expect it.

3. I neither like nor dislike it.
4. I dislike it.

*How would you feel if you did NOT have System/Tool 'X'?*
1. I would prefer not to be without it.
2. I could not work effectively without it.
3. I can manage without it, but might use it if it were still available.
4. I do not need it.

One other advantage of a simple questionnaire approach like this is that you can ask managers, or team/group leaders to answer the questions (as objectively as possible!) on behalf of their staff. Although not ideal, this enables a smaller population to be quizzed more quickly, yet still gives useful insights to the value of the software estate. For managers or team leaders, the questions then become:-

*How do you think the staff in your department/group/team feel about System/Tool 'X' now?*
1. They like it.
2. They expect it.
3. They neither like nor dislike it.
4. They dislike it.

*How do you think they would feel if they did NOT have System/Tool 'X'?*
1. They would prefer not to be without it.
2. They could not work effectively without it.
3. They can manage without it, but might use it if it were still available.
4. They do not need it.

In our modified Curlew-Kano Model, we still use the four original Kano categories: **Must-Be (M)**, **Performance (P)**, **Attractive (A)** and **Indifferent (I)**, and we plot the respondent's answers to the two questions according to the following grid in order to determine their "category of satisfaction" or CoS with the particular system or tool under consideration. Once you have a set of responses for the systems/tools under scrutiny (see Figure 3), you then move on to scoring those responses to ascertain whether that person is "satisfied" or not with each system. Then, when we have the category of satisfaction, we can assign a points' score for that system, for that person, as follows:-

- **M = Must-Be** (9 pts)
- **P = Performance** (6 pts)
- **A = Attractive** (3 pts)
- **I = Indifferent** (1 pt)

|  | Not like to be without | Cannot work effectively | Can manage without; might use | Don't need |
|---|---|---|---|---|
| **Like it** | P (6) | M (9) | A (3) | I (1) |
| **Expect it** | M (9) | M (9) | P (6) | I (1) |
| **Neither like nor dislike it** | P (6) | M (9) | A (3) | I (1) |
| **Dislike it** | I (1) | I (1) | I (1) | I (1) |

**Figure 3**: Category of Satisfaction: How a respondents answers pertaining to one system are plotted and scored.

So, for example, if Jane Doe, when asked about the LIMS system "Batch Tracker" says she "Likes it" and she "Cannot work effectively without it", then for her, this gets assigned a CoS of "Must-Be" and it scores 9 points. When John Smith is asked about the same system and responds that he "Neither likes nor dislikes it" and he "Doesn't need" it, then for him this system gets a CoS of "Indifferent" and it scores 1 point.

This example raises the obvious question around who gets asked about which systems? Clearly it's no good asking someone who works in HR about how satisfied they are with a clinical system and vice versa, but in this made up example, if John works in a group where "Batch Tracker" is used, then his low satisfaction with the system is very illuminating.

### Fun Fare

Whilst we cannot give real examples of how we have used our Curlew-Kano Model to help decision-making on potential system rationalisation, we can share a model example we really did try out with colleagues when trialling the approach.

We asked 5 co-workers about 6 apps on their mobile phones to see if we could identify, for these 5 guinea pigs, whether there were some apps, which they could seriously consider getting rid of. The six apps we chose were: (i) Camera; (ii) Social Media App; (iii) Map App; (iv) Taxi App; (v) Teleconferencing App (TC); (vi) Browser (please note, to save product embarrassment the actual app names are not given). The results of our initial survey are shown in Table 1.

It's important to note that not everyone answered the questions about every app, because some people said they couldn't comment on an app, which was on their phone, but which they hardly ever used. This raises another component that we had to consider for the bigger analysis: usage. We cover how to factor usage into the estate analysis in the next section in this article.

| App | Categories of Satisfaction [*# of Respondents*] | Total Curlew-Kano Score | Average [*Median*] Curlew-Kano Score | Priority | Initial Conclusion |
|---|---|---|---|---|---|
| Camera | 4 x M and 1x P [*5*] | (36+6)= 42 | 8.4 (*9*) | 2 | RETAIN |
| Social Media | 3 x I [*3*] | 3 | 1 [*1*] | 6 | REMOVE |
| Map | 2 x M and 1 x P and 2 x A [*5*] | (18+6+6)=30 | 6 [*6*] | 3 | REVIEW |
| Taxi | 3 x A [*3*] | 9 | 3 [*3*] | 5 | REMOVE |
| Tele-Conference | 1 x M and 2 x A and 1 x I [*4*] | (9+6+1)=16 | 4 [*3*] | 4 | REVIEW |
| Browser | 5 x M [*5*] | 45 | 9 [*9*] | 1 | RETAIN |

**Table 1**: Initial Survey Results

Our analysis allowed us to generate an important "score" which we could then use, in discussion with the stakeholders, to decide: which apps should be Retained; which could be Removed; and which might need to be Reviewed for retention or removal. This score we call the Curlew-Kano Satisfaction Score, or alternatively the System Satisfaction Score for each tool.

It should be noted that the Curlew-Kano satisfaction scores for a set of systems do not give the stakeholders or decision-makers an absolute guide to which systems should or should not be rationalised. So a total Satisfaction Score of 45 in the above analysis suggests "Retain"; but in another analysis of a different set of systems with a larger cohort of respondents, a score of 45 might suggest "Remove". It is important therefore to also consider the *average* (or for some investigations, the *median*) Curlew-Kano System Satisfaction Scores as these can provide the basis for a more normalised ordering and lead to better conclusions.

After normalisation, the Curlew-Kano Satisfaction Score can give an initial ordering or priority, where those with a high score might be deemed more worthy of retention compared to those with a low relative score. However, at this stage we must stress that no decisions on "Retain or Remove" should be made purely on the basis of just this score, and especially not when dealing with a small cohort of respondents. Why should this be so?

### Use your loaf!

We alluded earlier to the fact that it is vital to factor in system usage when considering rationalisation in order to get a full picture of the value of a particular system. Just using the System Satisfaction Score, even when normalised, would very likely lead to some serious rationalisation mistakes. For example, if a small cohort were asked about one particular system, which was in fact an important app for the business, and they happened not to be current, regular or frequent users of that system, the satisfaction score for that system might well be skewed to an artificially low level. Hence, as stated previously, gaining an insight to system or application usage, as well as asking the right people, is an absolutely critical second component for our software rationalisation methodology.

## 3. Factoring in usage

In most, if not all enterprises, it should be possible to get quantitative measures of how often a system is used and how many users there are for each system. However "usage" is a highly complex concept. Sometimes you have systems which are used by many people but only occasionally (e.g. expenses systems), and sometimes you have a system which is absolutely vital for a small business critical group and is used by a small number of people every day. So how did we accommodate usage in our approach?

**Use it or lose it?**

In the real-life project we were involved with, which led to our development of the Curlew-Kano Model, the timing was such that we could not get hold of reliable statistics that would give us the number of users and the amount of time a system was being used over a defined period. An added level of complexity also rests in the fact that the amount of time a system is "used" is not a simple concept. Some systems automatically log a user out after a defined period of inactivity, whereas others allow a user to stay logged in until they actively log out of the app or out of their PC (e.g. by shutting down at the end of the day). Also, a system may be used in bursts by individual users, so "usage" may fluctuate dramatically over, say, a 6 month period. This adds even more to the challenge surrounding objective quantification of "usage". In reality, for rationalisation projects of this type, lack of good, comparable usage data is likely to be the norm. To get around this deficiency, we added another question to our questionnaire in order to get a "category of usage", or CoU, assessment for each app directly from the users:-

*How often do you use System/Tool 'X'?*
1. L: A lot (every day or several times a week) = 4 points
2. S: Somewhat (once a week to once a month) = 3 points
3. O: Occasionally (2-4 times a year) = 2 points
4. N: Not very much or not at all (once a year or less) = 1 point

Whilst accepting that this is somewhat subjective, in our model survey, when we combined CoU with our Curlew-Kano Satisfaction Score, it allowed us to make some refinements in the rationalisation conclusions.

More specifically, when asked the category of usage for each app, our groups told us they used the phone apps as follows (see Table 2). For example, three respondents told us they used the Camera app a lot, and two said they used it somewhat. This allowed us to generate firstly a Total Usage Score and then a Usage Factor for each app, calculated by taking the Total Usage Score divided by the Number of Respondents.

| App | A. Category of Usage (CoU) Breakdown | B. Total CoU Score | C. Number of Respondents | Usage Factor (= B/C) |
|---|---|---|---|---|
| Camera | **3** x L and **2** x S | (12+6)=18 | 5 | 3.6 |
| Social Media (SM) | **1** x O and **2** x N | (2+2)=4 | 3 | 1.3 |
| Map | **1** x L and **4** x S | (4+12)=16 | 5 | 3.2 |
| Taxi | **3** x O | 6 | 3 | 2 |
| TC | **1** x S and **1** x O and **2** x N | (3+2+2)=7 | 4 | 1.8 |
| Browser | **5** x L | 20 | 5 | 4 |

**Table 2**: Model Survey results: Factoring in Usage

We then applied the Usage Factor to correct the Average Curlew-Kano Score and so give a more useful measure for making rationalisation decisions (see Table 3):-

| App | Average Curlew-Kano Score | Usage Factor | CK Score x Usage (CKU) | Priority | Conclusion |
|---|---|---|---|---|---|
| Camera | 8.4 | 3.6 | 30.2 | 2 | RETAIN |
| Social Media (SM) | 1 | 1.3 | 1.3 | 6 | REMOVE |
| Map | 6 | 3.2 | 19.2 | 3 | REVIEW |
| Taxi | 3 | 2 | 6 | 5 | REMOVE |
| TC | 4 | 1.8 | 7.2 | 4 | REMOVE |
| Browser | 9 | 4 | 36 | 1 | RETAIN |

**Table 3**: Combined Curlew-Kano and Usage Score

Factoring in usage with our average Curlew-Kano Satisfaction Score gave us what we are calling a Curlew-Kano-Usage or CKU Score, which allowed us to tease out a more granular rationalisation priority order and so allowed us to revise our initial conclusions/recommendations. In our small example, the Curlew-Kano-Usage Score indicated that the teleconferencing app was an even stronger candidate for removal.

## 4. Edge Cases

We now move on to a final factor which has the potential to be highly confounding when it comes to making informed and objective decisions about rationalisation: that factor is the tricky one of "edge cases".

In our initial analysis based just on the Curlew-Kano Satisfaction Score and not factoring in usage (i.e. NOT the CKU-Score), we suggested three broad initial conclusions, recommendations or outcomes for every system under scrutiny: **Retain**, **Remove** or **Review**. However, once you factor in usage, we believe there is a fourth 'R' which enables us to identify with more certainty the more tantalising edge case apps.

### Close to the Edge

In cases where there is only a very small cohort of responders, as in our model example, it is virtually impossible to identify apps which could be termed edge cases. Edge case systems tend to be outlier apps used by a small number of users (low Usage Factor), but which are nevertheless of high business value (High Satisfaction Score). We group these systems in a new category:"**Research**" – not because they are part of a Research or Discovery arm of an enterprise (though they often are!), but because they often require a much more informed, more deeply business-aware investigation to truly understand their value to the enterprise. In, for example, an R&D setting, edge case systems could include tools such as Pathology or Imaging apps.

Using a graphical presentation of our Curlew-Kano Satisfaction Score combined with the Usage Factor we can tease out this set of potentially difficult to deal with systems/applications. This ultimately gives us a 4R's mnemonic for software rationalisation: **Retain**; **Review**; **Remove**; **Research** (Figure 4):-

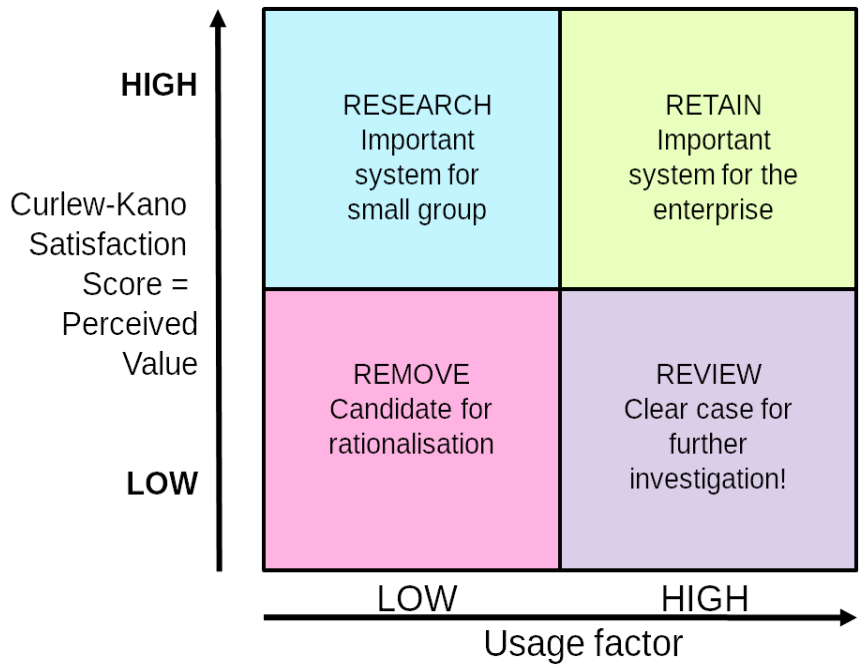

**Figure 4**: The 4R's mnemonic for system software rationalisation

But what do you do next once you have a Curlew-Kano-Usage Score and have assigned a system to one of the four 'R' categories?

## 5. What Happens Next?

It is perhaps important to stress at this stage that calculation of a combined Curlew-Kano-Usage (CKU) Score for every tool or system in an organisation's software estate cannot in isolation identify those tools or systems ripe for decommissioning. Every system will need some level of analysis or investigation before a decision can be made as to its longer term viability within the estate. However, an initial determination of the CKU Score can give the software rationalisation program a good, solid and objective indicator, by way of a ranking order, of those systems/tools which are likely to be the "low hanging fruit" at both ends of the spectrum - those which, with a little investigation, can be identified as: definite Retain systems; or definite Remove systems.

The main challenges in the rationalisation program will tend to sit with the remaining two system types: the ones that need Review; and the ones that need Research. The latter may be easier to come to a decision over as, by definition, they have a smaller user base, though a much deeper understanding of their use may be needed to ensure a truly objective decision on their future is reached. Despite this caveat, deeper diving into the Research, edge case systems' true usefulness and value should be relatively quicker and easier to achieve.

The bigger, potentially more resource-intensive challenge rests with the set of systems that are used by many people but whose satisfaction scores are at the lower level, thus leading to an intermediate Curlew-Kano-Usage Score and a Review categorisation. These are likely to be the tools that will require much more extensive, detailed, and hence potentially costly, investigation. Nevertheless, knowing early what those applications are, and how many there are of them can, at the very least, help the rationalisation program, especially its sponsor(s) and stakeholders, better understand the size and scope of the program's remaining activities.

So, our recommended blueprint for software rationalisation can be summarised as follows (Figure 5):-

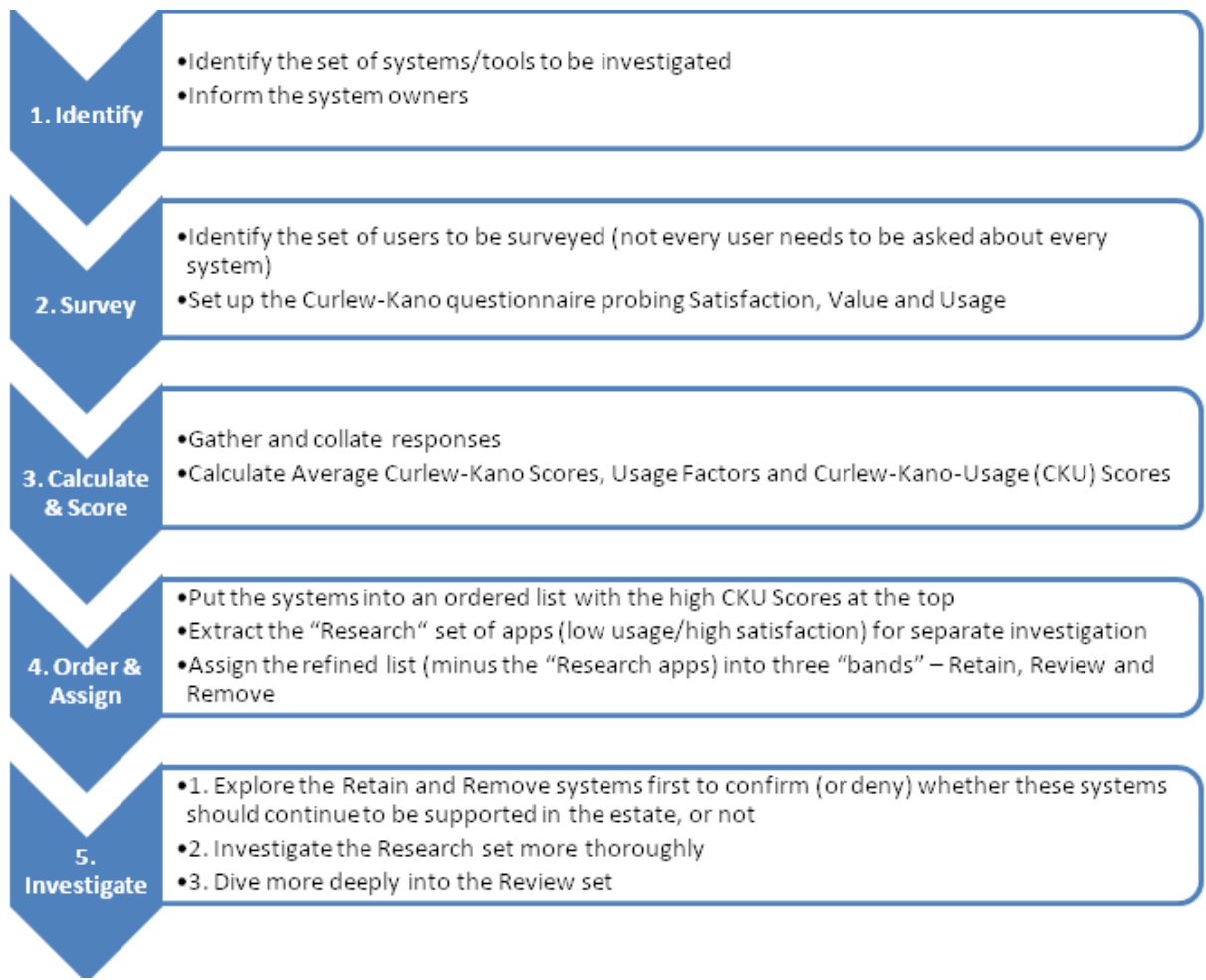

**Figure 5**: The Curlew Research blueprint process for software rationalisation

## 6. Conclusion

In the above discussion we present a method using our Curlew-Kano-Usage Score and a 4R's mnemonic for guiding software rationalisation programs towards making more informed and objective decisions about which systems are more likely to be prime contenders for retention or removal, and which may need more focused investigation. We also show how our methodology can be used to prioritise the rationalisation activities so as to target resources more effectively towards those systems, tools or apps that may need more intensive investigation.

It is worth acknowledging that there are many other approaches and methodologies available to help organisations with their software estate rationalisation. A selected few of these are given here (Apptio, 2022; LeanIx, 2022; Mackle, 2020; Orbus Software, 2022; Software AG, 2022).

There is even a "playbook" published by the CIO Council, which is a US Government forum of Federal Chief Information Officers (CIOs), with a six-step guide to application rationalisation (CIO Council, 2021).

The difference between our Curlew Research approach to those referenced above and others is that it is firmly focused on decommissioning from the perspective not of the IT department or the corporate business improvement group(s), but from that of the user. In our view it is the *user* who is best placed to judge a system, whether it is delivering value to them in their job, and even whether it is a tool they use at all. We do accept that directly asking people about their system likes and dislikes could introduce a risk of bias, whereby users who have a particular, possibly long standing attachment to a specific system could try and "game" the survey to ensure that their "pet" tool is not threatened. This risk of gaming the system is likely to be highest in the Research set of systems (low usage factor and high satisfaction score). But if the surveys are carried out honestly and objectively by all parties, and the users are assured that the goal of the Curlew-Kano methodology is not to finger systems for immediate decommissioning, but to order and prioritise the software estate so that follow-up investigation can be more effective, then we believe performing a survey such as we describe above can yield immensely useful information about the business value of the individual components of the enterprise's software estate.

Additionally, it is important to note that once you have a user-based "value proposition" for a system and a view on its priority and ranking within the enterprise's software estate, then a further investigations into: the application's "fit" with the corporate IT architecture and strategy; its total cost of ownership; whether it is close to the end or the "sunset"(Wierda, 2019) of its useful life - to name but three further important factors to be considered before a system can truly be identified for decommissioning - is a critical next step. But having the users' views upfront can provide an invaluable guide to whether a system's decommissioning is likely to constitute a good business decision. Running the Curlew-Kano Model survey will also inform users about the corporate rationalisation activity and so can, potentially, help to smooth the change program which will inevitably follow, once systems are formally identified at the enterprise level for rationalisation.

In conclusion therefore, we believe an objective, user-centric determination of satisfaction, value and usage by means of our Curlew-Kano-Usage (CKU) Score combined with our 4R's categorization provides a valuable insight to a system's true business value. Having a CKU Score and R-category for every system can help enormously in the decision-making that may need to take place at the enterprise level on software rationalisation and decommissioning.

Finally, it is important to stress that a system's CKU Score and R-category *will* change over time, so it is vital to perform a fresh Curlew-Kano-Usage analysis as part of each new round of software rationalisation. In summary, we believe that our simple, Kano-based, user-focused methodology can be implemented quickly and at relatively low cost by any organisation and provides a good, complementary approach that can sit alongside and enhance other application rationalisation methods. We believe using this method and involving users more fully in system rationalisation leads to better outcomes for the enterprise.